\documentclass[10pt,twocolumn,journal]{IEEEtran}

\usepackage{graphicx}
\usepackage{url}
\usepackage{placeins}
\usepackage{booktabs}
\usepackage{subfigure}
\usepackage{color}
\usepackage{amsmath}
\usepackage[first,light,timestamp]{draftcopy}

\begin{document}

\title{Secure Vehicular Communication Systems: \\ Implementation, Performance, and Research Challenges}

\author{F. Kargl$^{*}$, P. Papadimitratos$^{+}$, L. Buttyan$^{\dagger}$, M. Muter$^{\ddagger}$, E. Schoch$^{*}$, B. Wiedersheim$^{*}$\\
T.-V. Thong$^{\dagger}$, G. Calandriello$^{+}$, A. Held$^{\ddagger}$, A. Kung$^{\times}$, J.-P. Hubaux$^{+}$\\
\small
$^{*}$Institute of Media Informatics, Ulm University, Germany\\%
$^{+}$School of Computer and Communication Sciences, EPFL, Switzerland,\\%
$^{\dagger}$BUTE, Budapest, Hungary\\%
$^{\ddagger}$DaimlerChrysler AG, Research Vehicle IT and Services, Germany\\%
$^{\times}$Trialog, Paris, France%
}

\maketitle

\begin{abstract}
Vehicular Communication (VC) systems are on the verge of practical deployment. Nonetheless, their security and privacy protection is one of the problems that have been tackled only recently. In order to show the feasibility of secure VC, implementations are required. In \cite{PapadBH:08} we have discussed the design of a VC security system that has emerged as a result of the European SeVeCom project. In this second paper, we discuss various issues related to the implementation and deployment aspects of secure VC systems. Moreover, we provide an outlook on open security research issues, which will arise as VC systems develop from today's simple prototypes to full-fledged systems.
\end{abstract}

\section{Introduction}

Vehicular communication (VC) systems will enable many exciting applications that will make driving safer, more efficient and more comfortable. But this necessitates the introduction of security and privacy enhancing mechanisms, as discussed in~\cite{PapadBH:08}. In this paper we focus on practical aspects associated with the implementation and deployment of such a secure VC system. We also provide an outlook to future research challenges.

First, we motivate why the deployment of a security system for a vehicular environment is different compared to other common information technology systems. We then present the SeVeCom baseline architecture, and highlight various implementation- and deployment-specific aspects such as: flexible integration in existing communication stacks, use of a hardware security module, and secure connections of VC on-board units to in-vehicle bus systems. Furthermore, we analyze performance and communication overhead of the suggested security mechanisms and propose optimizations for efficient secure communication.

Finally, we present selected topics we consider relevant for future research on VC system security. One aspect is the use of complex forms of data dissemination, such as aggregation schemes, which require different security approaches than the ones used for broadcast and unicast communications. Another aspect is the integration of VC systems with other networks or connecting them with mobile commodity devices, which raise additional security problems. Other future research aspects include secure localization and the question whether existing VC privacy solutions are indeed sufficient.


\section{Vehicular Communication Systems}\label{sec:svc}

There are significant differences between devices such as mobile phones or desktop computers connected to the Internet and devices in a VC system. Differences in development, production, and operation, determine VC-specific constraints and conditions:

\begin{enumerate}
\item Vehicles have a long life span, lasting several years in most cases. This makes it hard to change on-board systems as reaction to new upcoming risks to the vehicle safety.

\item Owners have constant physical access to and full control over vehicles. In spite of the involved safety risks, many users might try to modify or ``enhance'' their vehicles. From a manufacturer's point of view, the risk of hardware tampering cannot be neglected. 

\item No technical expertise on vehicle electronics or VC security aspects is expected from a user that runs a vehicle. Hence, the vehicular security measures have to operate autonomously with no need for intervention or feedback from the user.

\item Robustness requirements and time constraints are demanding. Functions necessary, for example, for driving or alerts received via the VC system must be processed in real-time: delays or errors could lead to vehicle malfunctions, driving errors, and consequently to physical damages and injuries.

\item Liability and conformance require precise formulation of legal issues. Differing regulations and requirements in various countries make it even more difficult to address these challenges.
\end{enumerate}

These observations have consequences on the implementation of a VC security system. Due to the long vehicle life cycle, it cannot be ensured that all threats are thwarted at the time of development. Therefore, the VC security mechanisms should be flexible, adaptable, and extensible, to allow later adjustments to changing security requirements. To address this need, we propose a component-based security architecture for VC systems, which allows to add, replace, and reconfigure components (for example, substitute cryptographic algorithms) throughout the life cycle of the vehicle.

The large number and the variety of vehicles have to be taken into account. Even for a single car type, different production and equipment lines lead to many distinct versions and variants. Nonetheless, it should be possible to integrate a security system into all those platforms. In addition, the communication stack and security measures might be designed by different teams or vendors; a situation that clearly requires well-defined but still flexible interfaces. These reasons led to the development of the so called ``hooking architecture'', which introduces special hooks at the interface between every layer of the vehicular communication system. The hooking architecture introduces an event-callback mechanism into the communication stack which allows adding security measures without the need to change the entire communication system. 
The security system in a vehicle has to fulfill real-time or near real-time requirements. For the underlying cryptographic primitives, this implies optimized cryptographic hardware, in order to guarantee the near real-time performance. The potential trade-off between security and performance has to be well balanced.

To enable VC systems to withstand future, yet unknown attacks, besides the traditional prevention-oriented approach, functionalities to detect attacks, such as intrusion detection capabilities, and to recover after an attack, are needed. In the long run, the goal is to enhance the resilience of the system.

\section{SeVeCom Implementation}

The SeVeCom project defines a baseline security architecture for VC systems~\cite{deliverable2.1}. Based on a set of design principles, SeVeCom defines an architecture that comprises different modules, each addressing certain security and privacy aspects. Modules contain components implementing one part of system functionality. The baseline specification provides one instantiation of the baseline architecture, building on well-established mechanisms and cryptographic primitives, thus being easy to implement and to deploy in upcoming VC systems.

\subsection{Baseline Architecture: Deployment View}

The SeVeCom baseline architecture addresses different aspects, such as secure communication protocols, privacy protection, and in-vehicle security. As the design and development of VC protocols, system architectures, and security mechanisms is an ongoing process, only few parts of the overall system are yet finished or standardized. As a result, a VC security system cannot be based on a fixed platform but instead has to be flexible, with the possibility to adapt to future VC applications or new VC technologies.

To achieve the required flexibility, the SeVeCom baseline architecture consists of modules, which are responsible for a certain system aspect, such as identity management. The modules, in turn, are composed of multiple components each handling a specific task. For instance, the Secure Communication Module is responsible for implementing protocols for secure communication and consists of several components, each of them implementing a single protocol. Components are instantiated only when their use is required by certain applications, and they use well-defined interfaces to communicate with other components. Thus, they can be exchanged by more recent versions, without other modules being affected.

\begin{figure}[t]
\begin{center}
\includegraphics[width = \columnwidth]{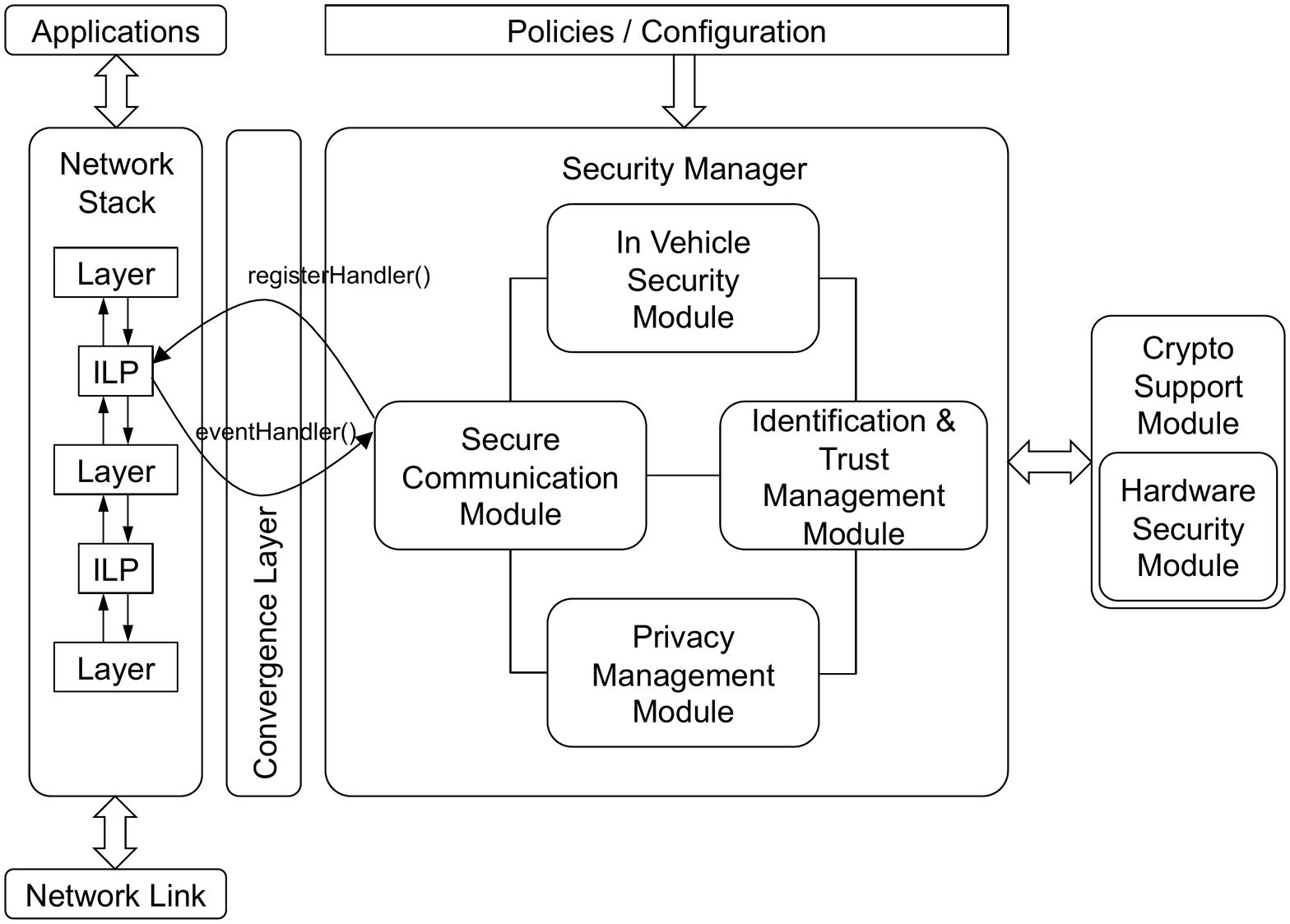}
\caption{Baseline Architecture: Deployment View.} \label{fig:bl-arch}
\end{center}
\end{figure}

As shown in Fig.~\ref{fig:bl-arch}, the \emph{Security Manager} is the central part of the SeVeCom system architecture. It instantiates and configures the components of all other security modules and establishes the connection to the Cryptographic Support Module. To cope with different situations, the Security Manager maintains different policy sets. Policies can enable or disable some of the components or adjust their configuration, for example, to enhance or relax the parameters for a pseudonym change under certain circumstances.

\subsection{Communication Stack Integration}

To be independent of the actual communication stack, the integration of the SeVeCom security system into the protocol stack is based on a \emph{hooking concept}, inspired by similar architectures such as the Linux Netfilter kernel subsystem. \emph{Inter Layer Proxies} (ILPs) are inserted at several points in the communication stack. Every ILP maintains a list of callback handlers that are to be notified of certain events.

During initialization, the SeVeCom components can register at an ILP, subscribing for certain message types and direction (up or down the stack). Therefore, they have to implement an event listener interface and use the \emph{registerHandler()} method to connect to an ILP. Some components may have to register at multiple ILPs, subscribing for different kinds of packets. When a message arrives at an ILP, an event callback is triggered for all components that have registered for this message type and their \emph{eventHander()} method is called. The callback includes a reference to the received message, and the component is then able to inspect or modify it. By the return value the component indicates if the message was modified, if it should be reinserted into the stack, or if it should be simply dropped by the ILP. The Secure Beaconing Component, for example, connects to the ILP above the MAC layer and checks the signatures of all incoming beacon messages. Beacons with invalid signatures are either discarded or tagged. Using this hooking architecture, it is possible to transparently integrate security functionality into an existing network stack with minimal modifications.
Whereas events are triggered by the communication stack, the security system can also access the stack by means of command calls using a well-defined API offered by stack layers. Command calls could, e.g.\ instruct the MAC layer to set its MAC address to that of a new pseudonym.

The hooking concept makes certain assumptions about the network stack. It assumes a layered architecture, where the ILPs can be inserted in between, and the stack has to implement a certain command API, e.g.\ for change of MAC addresses. To be able to port the SeVeCom architecture to many different communication platforms, we also provide an additional convergence layer: This defines an abstraction interface that proxies call between the communication system and the security components. Whenever the SeVeCom system is ported to a new platform, besides adapting to different packet formats, only the ILPs and the convergence layer have to be modified, while all other components remain unaffected both in terms of security and communication.

\subsection{Hardware Security Module}

As explained in~\cite{PapadBH:08}, the purpose of the Hardware Security Module (HSM) is to provide a physically protected environment for the storage of private keys and for the execution of cryptographic operations using them. Clearly, the full implementation of a HSM is beyond the scope of the SeVeCom Project, but we can summarize the main requirements that such an implementation should meet in order to be applicable for securing vehicle communication systems.

First of all, the HSM must be tamper resistant, to some extent. High-end tamper resistant modules (such as the IBM 4758 Cryptographic Coprocessor) are too expensive to be added to every vehicle. At the same time, we observe that low-end tamper resistant devices (such as smart cards) do not provide all the functionality that we need. In particular, commercially available low-end devices do not have built-in batteries, and consequently, cannot provide a trusted internal clock. As pointed out in~\cite{sevecom2007}, without a trusted source of time, such devices are not able to produce time-stamps that can be trusted by other participants of the system. Therefore, we need an HSM implementation somewhere between high-end and low-end devices. A potential approach is to implement the HSM as an Application-Specific Integrated Circuit (ASIC) with some special coating that provides a certain level of tamper resistance. Such a customized device can provide all the necessary functionality by design and it can be produced in large quantity at sufficiently low costs.

Second, the HSM must have an API, through which it can provide services to the other modules of the security architecture that run on the OBU. This API should support the digital signature and timestamping service, the decryption service, as well as the key and device management services described in~\cite{PapadBH:08}. We specified such an API in the SeVeCom Project, however, lacking the appropriate HSM hardware, we only implemented it in the form of a software library running on a general purpose computer. Nevertheless, besides being useful for demonstration purposes, our implementation can also serve as a reference for future implementations on real HSM devices. In our implementation, we used ECDSA for digital signature generation, ECIES with HMAC-SHA1 and AES-CBC for encryption, and we fully implemented the key management services of the HSM described in~\cite{PapadBH:08}.

Finally, we note that some examples published in~\cite{BondA:01} show that physically secure modules can successfully be attacked through their weakly designed API. For this reason, we used formal verification techniques to verify the SeVeCom HSM API. Our method is based on the applied pi-calculus and an automated verification tool called \emph{ProVerif}. We proved that a key generated by an adversary cannot be implanted as a new root key in the HSM through the API. Additionally, short-term and long-term private keys are proved not to be revealed as the result of possible series of function calls.

\subsection{In-Vehicle Security}

In order to achieve their full potential, VC systems need access to the in-car network and sensors that observe the current status of the vehicle and the environment. This enables the VC system to process signals such as emergency braking, airbag activation, and slippery road detection, thus greatly contributing to the avoidance of accidents and improvement of road safety.

On-board system signals are transferred inside the car through different networks and domains. Usually, the network architecture and the in-car gateways restrict the signals to the defined network segments and prevent information from leaving its dedicated domains. This clear architecture and strict separation is one measure that ensures the entire vehicle, especially its vital functions (brakes, engine or airbag control), always operate reliably and \emph{cannot} be attacked from the outside. If this were to be changed into a more open architecture, for example, allowing reading out sensor information from in-vehicle networks or displaying and reacting to warning messages from external sources, it would be absolutely necessary to ensure that in-vehicle systems are protected from any external malicious influence.

The \emph{In-vehicle Security Module} protects the interface between the in-car networks and the wireless communication system. It controls external access to the in-car networks, on-board control units and vehicle sensor data, but it also ensures that data and services required by other V2V and V2I applications are provided correctly. Within the in-vehicle security module, two main components are provided: (i) A \emph{firewall} that controls the data flow from external applications to the vehicle and backwards, and (ii) an \emph{Intrusion Detection System} (IDS) that constantly monitors the status of the in-car systems and provides real-time detection of attacks.

The firewall realizes a packet or application based firewall approach. Its rule-based table states which application is allowed to access each kind of data or service. The IDS can dynamically add rules to the firewall table, in order to deny access for a specific application or disable a service.

The IDS is based on an anomaly detection approach, which implies that normal on-board system behavior is clearly defined and specified. If an event results in an on-board system state that is not part of the standard specification, a potentially dangerous situation is detected. Depending on the source and type of the event, appropriate reactions are taken to get the system back to a secure and safe state.

\section{Performance Issues}

One very important aspect towards deployment is performance. Given cost constraints in today's car manufacturing, one cannot equip vehicles with state-of-the-art desktop processors. Instead, cheap and energy-saving embedded processors are used. At the same time, cryptographic operations to secure VC\cite{PapadBH:08} create a significant overhead both in terms of processing and communication bandwidth.

This is especially true because beaconing is a fundamental VC protocol: Vehicles frequently send information (e.g., position and environment conditions), typically one beacon per 100 milliseconds. At these rates, the security overhead will be significant. Without ignoring other factors, the computational security overhead is due to generation and verification of packet signatures and certificates. The communication security overhead is due to signatures and certificates attached to packets. Each safety beacon has to be signed, and each vehicle has to validate, for example, every 100 milliseconds, beacons from \emph{all} neighboring vehicles in range, which, not to forget, may also change their identity (pseudonym) in the meantime.


While RSA and DSA signatures have long been industry standards, these mechanisms do not meet the overhead requirements both in terms of processing and bandwidth. Especially in combination with large X.509v3 certificates, they are unsuitable for high-speed and low-overhead VC systems. In contrast, for the same security levels Elliptic Curve Cryptography (ECC), that is, ECC signatures, keys, and certificates, are significantly smaller than their RSA and DSA counterparts. This is the reason why SeVeCom as well the IEEE 1609.2 trial standard chose to utilize EC-DSA signatures. In addition, SeVeCom utilizes compact certificates.

To reduce overhead,~\cite{CalaPHL:07,PapadCLH:08} propose to \emph{not} attach certificates to all messages, but rather do that for one every $\alpha$ successive beacons; they also propose certificate caching to reduce verification processing overhead. Additional optimizations are proposed in~\cite{KarglSW:08}: Omitting signatures or signature verifications in certain situations, and avoiding attaching certificates based on the context, that is, \emph{unless} a change in the vehicle neighborhood takes place.

Such overhead can affect VC applications in multiple ways. An investigation on safety applications is provided by~\cite{CalaPHL:07,PapadCLH:08}. The first dimension of the problem is communication reliability: increased beacon size contributes to interference. In principle, the higher the offered load, with the number of transmitters in the area, the beaconing rate, and the message overhead, the worse the channel performance.

The second dimension is processing overhead: Each receiver V must in principle verify a signature for each received packet, while signature generation is not as critical (in general, V signs one and verifies N messages per time slot). Simulations in~\cite{PapadCLH:08} show that V's CPU is heavily stressed in situations with dense topologies, for example, in congested multi-lane highways, even if the vehicle direction is used to avoid processing messages from vehicles on the opposite flow.

These findings assume hardware for on-board units that are used for current VC prototypes; in upcoming field-trials OBUs are expected to have less powerful hardware like a Power PC CPU at 400 MHz~\cite{Tim:07}. Initial products presumably will be equipped similarly. Actual crypto performance of this hardware depends very much on implementation (e.g. if pre-calculated tables are used), but assuming efficient software libraries and ECDSA-224, we estimate that this hardware will not be able to process more than a few dozens of verifications per second. Dedicated ASICs are expected to be able to handle the required cryptographic load at moderate costs~\cite{Baktir:08}.

By integrating optimization mechanisms to pseudonymous authentication, as those of\cite{CalaPHL:07,PapadCLH:08,KarglSW:08}, we introduce additional verification delays: Safety warnings can be trusted only if the corresponding short-term certified public key (pseudonym) was previously verified. It turns out that the overall impact of security on this front can be kept low, with the number of crashes experienced in a platoon of vehicles is close to that achieved without any security mechanism~\cite{CalaPHL:07,PapadCLH:08}. An additional mechanism of repeatedly attaching a certificate to $\beta$ successive beacons when a pseudonym change takes place can increase reliability. We note however that the performance of safety applications is heavily influenced by other parameters, like the placement of vehicles, the beaconing rate and the penetration rate of vehicular communication.

\begin{figure}[t]
\begin{center}
\includegraphics[width = 0.95\columnwidth]{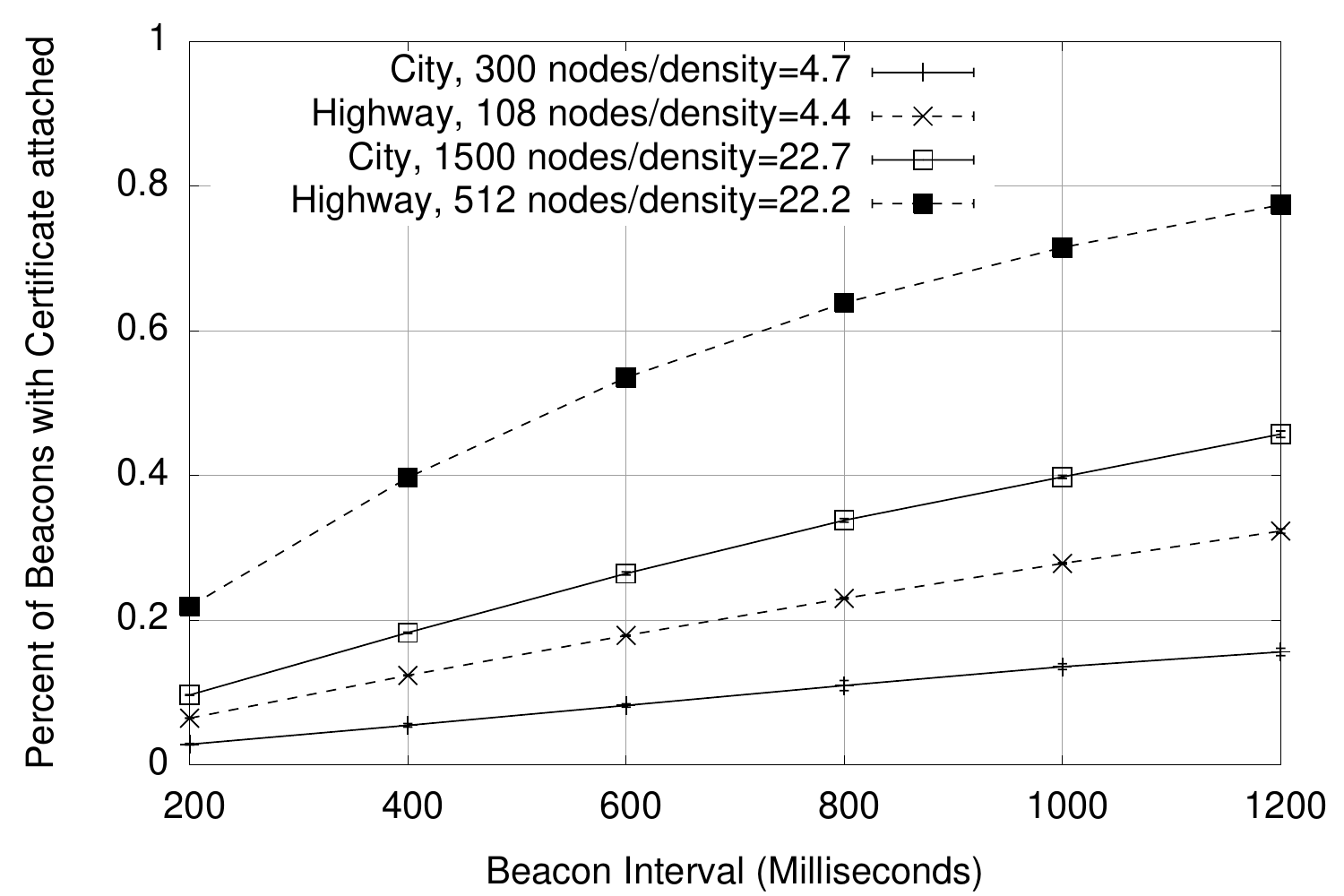}
\caption{Performance Results: Beacon Certificate Omissions.}\label{fig:beacon}
\end{center}
\end{figure}

\begin{figure}[t]
\begin{center}
\includegraphics[width = \columnwidth]{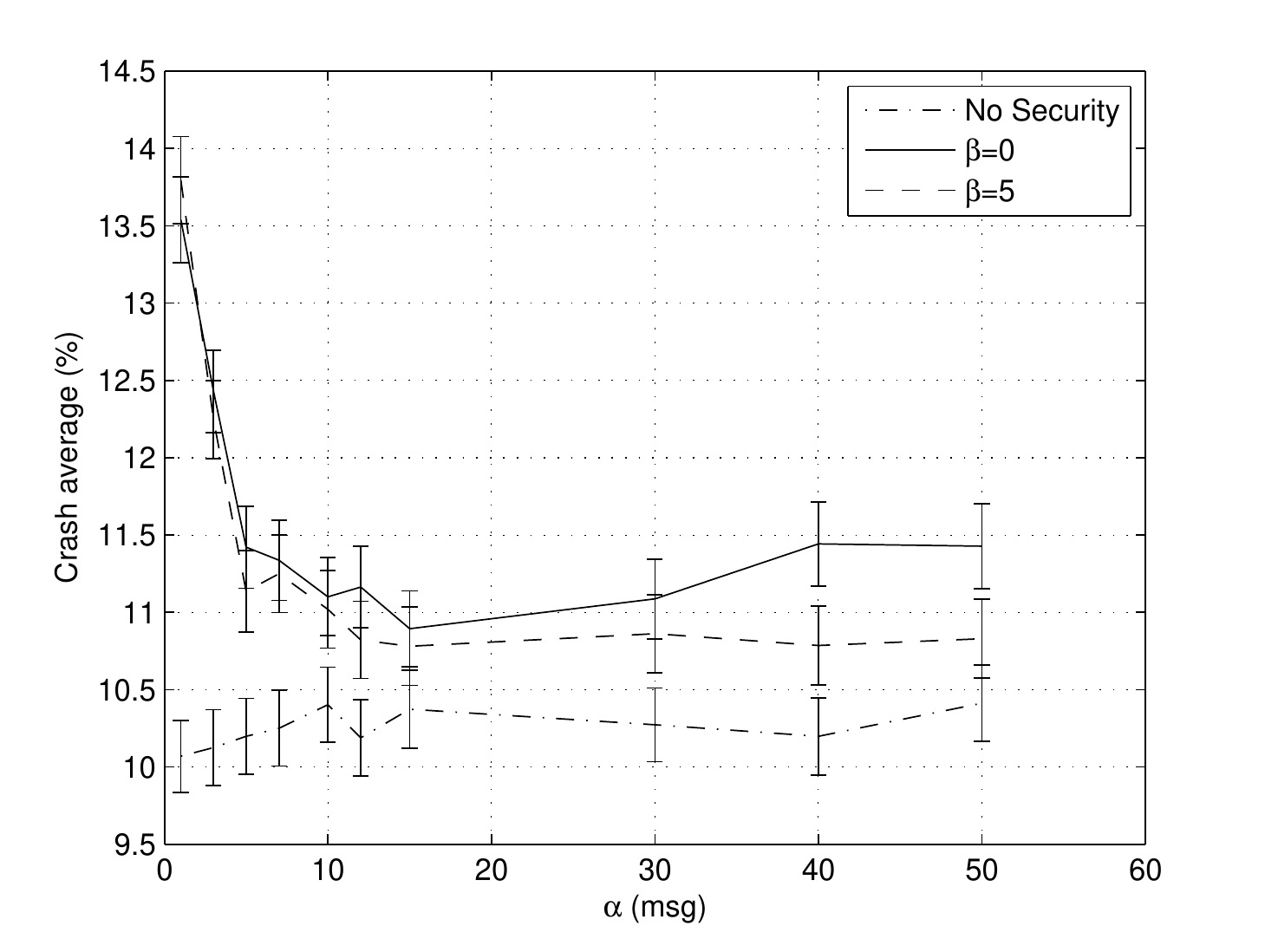}
\caption{Performance Results: Impact of Security and Privacy on Safety.} \label{fig:safety}
\end{center}
\end{figure}

Based on the optimizations discussed above, Fig.~\ref{fig:beacon} shows the fraction of beacon messages with attached certificates in various scenarios and several beaconing intervals. Certificates are attached to only beacons if new neighbors are discovered. The result is that certificates can be omitted in more than 90\% of all sent beacons for small beacon intervals and medium node density. Fig.~\ref{fig:safety} shows the impact of security on safety for an emergency braking notification application in a challenging, dense fast-moving network: with $1/\alpha$ the fraction of omitted certificates and $\beta$ the number of repetitions upon pseudonym change, the effectiveness of unsecured VC is practically the same as that of secure VC.

\section{Research Challenges}

We consider our VC security solution, described here and in~\cite{PapadBH:08}, as mature and practically deployable. Nonetheless, there are open issues that cannot be handled by existing security strategies alone, thus calling for new approaches. We highlight those, present initial results, and raise questions towards future research.

\subsection{Alternative Communication Forms}

On-going research has mostly considered VC protocols relying on periodic beaconing, flooding, GeoCast, and position-based routing. Up to now, these mechanisms have received the attention of work on VC security and privacy. Nonetheless, recently, additional means of information dissemination have been considered in the context of VC. For example, the literature highlights the need for more efficient flooding and GeoCast strategies, and suggests the use of Gossiping or Context-adaptive Message Dissemination, as well as Data Aggregation in VC systems~\cite{SchochKLW:08}.

These new approaches will necessitate an adaptation of security and privacy strategies. Mechanisms such as Context-adaptive Message Dissemination already provide an inherent degree of resistance against attacks \cite{SchochKKW:08}. In contrast to many routing protocols, where the protocol itself can become the target of an attacker, there is (nearly) no signaling between nodes that an attacker could exploit.

Another aspect that brings forth new security issues is the need for nodes that relay messages to also modify them. This has been already the case for position-based forwarding we discussed in \cite{PapadBH:08}. But this is even more so for Context-adaptive Message Dissemination and Data Aggregation. In the latter case, individual information contributed by vehicles usually becomes unavailable during the dissemination process.

Misbehavior, including injection of erroneous data and denial of service (DoS), is still possible, even if strong cryptographic security is present. Mechanisms that perform consistency checks, using redundant information or on-board sensors, can be used to discard incorrect information from the network. Meanwhile, rate limits can confine the effects of DoS attacks. Initial exploration of such mechanisms has already delivered promising results \cite{SchochKKW:08}.

\subsection{Data-centric Trust}

We observe that the trustworthiness of messages sent by a node (vehicle or RSU) is primarily determined by the trustworthiness of the sender's credentials. Essentially, the VC system entities, CAs and nodes, make statements on public keys, identities, and attributes, and data and VC messages respectively. Then, at any point in time, messages from any newly encountered car are trusted as long as its certificate is valid. Such trust relations, entity-centric and set \emph{a priori}, are useful, but they lack the flexibility that is necessary for highly volatile and data-centric VC systems.

Given the majority of VC applications, it becomes clear that it is often more useful to assess the trustworthiness of data \emph{per se}, rather than assessing only the trustworthiness of the nodes that report them. The need for \emph{data-centric trust establishment} is clearer if we consider that identities of nodes are largely irrelevant, even if no privacy enhancing mechanisms are employed. In contrast, it is the data (e.g., safety warnings, traffic information) and their freshness and location relevance that matter the most. From another point of view, trying to interact with possibly adversarial (faulty) data senders to determine their trustworthiness is hard: encounters are in general short-lived and with no prior association.

Considering more generally the issue of non-cryptographic protection mechanisms, it is possible to rely on own or trusted measurements, for example, as discussed in \cite{PapadBH:08} for securing GeoCast, or in the previous section for context-aware data dissemination, to discard erroneous data. But data may come from relatively remote sources. More important, the receiving node will be unable to determine their trustworthiness alone. So a cooperative management of data-centric trust is needed. Beyond what is presented in~\cite{infocom08}, further development of techniques to achieve data-centric trust establishment will be needed.

\subsection{Secure Localization}

Location information is critical for VC systems, especially for cooperative awareness, collision avoidance, and essentially all safety applications, as well as for position-based information dissemination. An internal adversary could announce false own positions, while an input-controlling adversary~\cite{PapadBH:08} could affect the position announced by its victims, and this way disrupt or abuse the VC operation. Whereas an internal adversary could be thwarted by data consistency checking and position verification, these methods cannot be effective against an input-controlling attacker that attacks the Global Navigation Satellite Systems (GNSS).

The objective of the adversary is to manipulate the location GNSS receivers compute, for example, for the Global Positioning System (GPS). To do so, the adversary can interfere with GNSS transmissions and inject forged navigation messages. A variant of such attacks, termed replay attacks, is possible even if GNSS were cryptographically protected. In fact, replay attacks can be fine-grained, so that gradual manipulation of each victim location can remain small and thus hard to detect. But, cumulatively, they can lead to substantial distances between the actual and the perceived (provided by the GNSS) location of the victim nodes~\cite{iwssc08}. Equally interesting, such attacks are possible without any compromise, physical or not, of the GNSS receiver or other on-board equipment or software.

This leads to an important realization: location information in the system cannot be considered by default trustworthy. One solution would be to leverage on mechanisms as the secure neighbor discovery and position verification discussed in~\cite{PapadBH:08}. In addition, dedicated infrastructure could provide ``land-marks,'' assisting the detection of false location information. Finally, mechanisms that detect adversarial GNSS transmissions could be devised and integrated in the GNSS receivers or the OBUs. If so, correct nodes falling prey to an input-controlling adversary would declare their own location information as faulty, and thus refrain disseminating in any messages they transmit. We believe that future efforts in those two main directions should be undertaken.

\subsection{Secure Integration of Commodity Devices}

Devices such as portable computers, mobile phones, iPods, or (portable) navigation systems are becoming more and more ubiquitous. Customers wish to use these devices inside of the vehicle, and connect them to the vehicle electronics where meaningful. Nowadays, mobile phones and iPods can already be connected to vehicles up to some degree. In the future, complete and seamless integration is desired.

Portable navigation systems, for instance, could be improved by transferring data from the vehicle's rotation sensors of the wheels and the current velocity to improve navigation in tunnels. For the calculation of the route and the arrival time, additional internal data, such as fuel status, could be taken into account. If the customer were allowed to connect her mobile computer to the vehicle network in an uncontrolled way, she could be given the chance to check the vehicle status in detail, and change at will settings like the engine configuration or the visual layout of the telematic system user interface.

Every interface and connection of non-VC devices to the in-vehicle system poses a threat and increases the risk that malicious code or adversaries gain access to the in-vehicle system. Wireless interfaces raise additional concerns, as illegitimate access could be easier and achieved from a distance. To prevent in-vehicle and thus VC system compromise, it is necessary to define specific policies that describe and devise security mechanisms that enforce parsimonious access of commodity devices to in-vehicle resources.

\subsection{Hybrid Vehicle Communication Systems}

\begin{figure}[t]
\begin{center}
\includegraphics[width = \columnwidth]{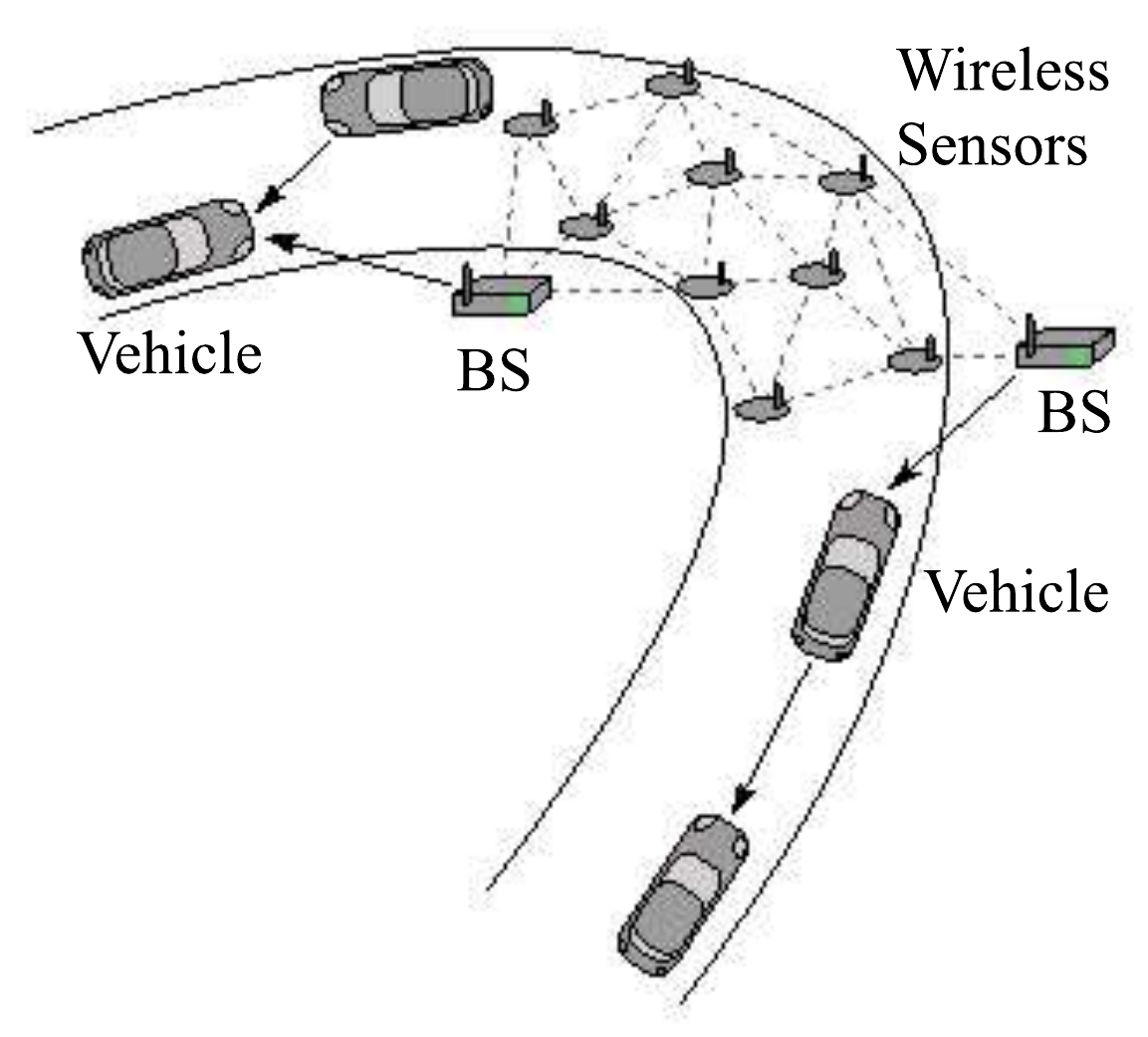}
\caption{Hybrid Vehicle Communication System.} \label{fig:hybrid}
\end{center}
\end{figure}

VC systems could be integrated with other communication networks, such as cellular networks, WiFi networks, wireless sensor networks, and mesh networks. They could, for instance, take advantage of the ubiquitous coverage provided by cellular networks, especially in the initial deployment phase when their penetration rate is expected to be low. Beyond the obvious use of cellular data services for information download - including security related data - the cellular infrastructure could also be used for geo-casting traffic and safety related information with less stringent delay requirements; systems in this direction have been investigated for example by the European Com2React project~\cite{com2react}. WiFi networks could be used for similar purposes - at higher data rates - in urban areas where WiFi coverage is substantial. Wireless sensor networks deployed along hazardous roads can collect and process local environmental information and share that with vehicles passing by. Fig.~\ref{fig:hybrid} shows an example of such a hybrid VC system that also incorporates sensor network nodes which deliver sensing data via a base station to nearby vehicles.

In terms of security, the integration of vehicle communication systems with other communication networks requires at least an integrated authentication infrastructure. In particular, vehicles need to authenticate the messages that they receive from cellular base stations, WiFi access points, and roadside sensors before trusting and acting upon them.
While the integration with cellular networks seems to be straightforward, by adopting the well-established security mechanisms of those networks, the integration with WiFi and sensor networks is more challenging. One problem is roaming across WiFi access points, because different operators may support different authentication methods, and even if a common method is accepted (e.g., based on the TLS Handshake Protocol), a large scale, international PKI needs to be available. In case of wireless sensor networks (WSN), the challenges bear similarities to those for VC systems, with additional problems stemming from sensor node resource constraints and lack of physical protection. Even if sensor nodes could authenticate themselves to vehicles, they could have been tampered with and compromised, and therefore their data may not be trusted. In fact, WiFi and WSN operators may not be trusted to the same extent as cellular network operators, for not misusing (e.g., sharing) sensitive for the users' privacy transactional data.

\subsection{Privacy}

Taking into account privacy concerns when designing vehicle communication systems is important for several reasons. First of all, privacy is a basic right of people, and we believe that new technologies should be designed in such a way that they make it possible to retain this right. In addition, the protection of privacy is made mandatory by laws in many developed countries.

For these reasons, we integrated a baseline privacy protection mechanism into our architecture based on using and changing pseudonyms. However, the proposed pseudonym mechanism has some limitations: It might still be possible to fully track vehicles between pseudonym changes. Increasing the frequency of changes can help, but it also increases the incurred overhead. In addition, taking into account statistical models of the traffic in a given geographical area, tracking of vehicles is possible to some extent despite frequent pseudonym changing and despite the potentially limited observational capabilities of the adversary, as discussed in~\cite{Buttyan2007a} and~\cite{PapadBH:08}.

Hence, there is a need for new and improved privacy protecting mechanisms that provide stronger guarantees. One promising approach is based on group signatures, however, the efficiency of those signature schemes must be substantially increased before they can be deployed in practice. In the meantime, hybrid solutions can be envisioned such as the one proposed in~\cite{CalaPHL:07}.

Moreover, attacks against privacy may happen at any layer of the communication stack. So far, most of the research efforts have focused on privacy enhancing technologies in the Medium Access Control (MAC) layer and above. However, recent advances in radio fingerprinting techniques make it possible to identify an RF device at the physical layer. Unfortunately, attacks at the physical layer may render protection at higher layers ineffective. Therefore, some research effort is needed to address the problem of using radio fingerprinting for tracking purposes.


\section{Conclusions}


Securing vehicular communication (VC) systems is complex endeavor, with multiple facets and subject to several unique constraints. We have systematically analyzed the problem at hand, identifying pertinent threats and models for adversaries. We considered general security requirements, and mapped those to specific VC applications. Based on a set of design principles, aiming at a practical system that can be readily adopted towards deployment, we designed a comprehensive solution, a security architecture for VC systems. We focused on identity and credentials management, security for a variety of communication protocols, and privacy enhancing mechanisms. Furthermore, we proceeded with experimental evaluations of our mechanisms, based on simulations and prototype implementations. Our results show that with the appropriate design secure VC systems can be practical, able to support VC applications as effectively as unsecured VC systems would. Moreover, our security architecture implementation could be ported with minimal modifications to practically any platform. With the SeVeCom project reaching its conclusion, we identified and made progress towards addressing additional research questions. This is why we believe our system can be the basis for the deployment of robust, user privacy-preserving, secure VC systems. 

\bibliographystyle{plain}
\bibliography{refs}

\end{document}